# Structuring Spreadsheets with the "Lish" Data Model

Alan Hall, Michel Wermelinger, Tony Hirst and Santi Phithakkitnukoon.
The Open University, UK and (4th author) Chiang Mai University, Thailand.
alan.hall@open.ac.uk

## ABSTRACT

*A spreadsheet is remarkably flexible in representing various forms of structured data, but the individual cells have no knowledge of the larger structures of which they may form a part. This can hamper comprehension and increase formula replication, increasing the risk of error on both scores. We explore a novel data model (called the "lish") that could form an alternative to the traditional grid in a spreadsheet-like environment. Its aim is to capture some of these higher structures while preserving the simplicity that makes a spreadsheet so attractive. It is based on cells organised into nested lists, in each of which the user may optionally employ a template to prototype repeating structures. These template elements can be likened to the marginal "cells" in the borders of a traditional worksheet, but are proper members of the sheet and may themselves contain internal structure. A small demonstration application shows the "lish" in operation.*

## 1. INTRODUCTION

Building a spreadsheet frequently involves a high degree of replication, both at the level of cells containing the same or similar formulae and at the level of higher structures such as families of similar tables. In software engineering terms this is a contravention of the "don't repeat yourself", or DRY, principle [Thomas & Hunt, 1999]. It increases the risk of errors due to possible inconsistency among the repeated elements, and can make maintenance particularly problematic.

There is a long history of developments aimed at capturing some of the higher level structures to be found in spreadsheet models, reducing the need for replication. Notable early efforts were Lotus Improv (in 1991), which introduced column formulae and named dimensions; and the pioneering work of [Burnett et al., 2001] whose "dynamic grids" allowed tables (and regions of them) to be manipulated as discrete objects. Excelsior [Ireson-Paine, 2005] banished the replication entirely, at the expense of requiring the user to write code in a spreadsheet description language. More mainstream examples include the availability of array formulae, pivot tables, and ranges designated as "Tables" in Excel.

One potential barrier to uptake of more structured approaches like these is that they require the user to master new abstractions over and above the engagingly simple one that is the standard spreadsheet grid. Our approach is therefore based on a modification of the grid itself – a new abstraction of a kind, to be sure, but a rather simple one – as opposed to any new apparatus to be supplied alongside it. We borrow from [Erwig et al., 2006] the notion of a template that can generate a structure, but do so on-sheet and with no prior assumptions about what forms of table are legal.

Recently this area has seen a flurry of activity, partly arising from the application of spreadsheet-like approaches to relational database queries [McCutchen et al., 2016;

Bakke & Karger, 2016; Chang & Myers, 2016]. We place our work more squarely in the domain of general purpose spreadsheets, along with [Hodnigg & Pinzger, 2015] who developed procedures for identifying areas of the worksheet by role and similarity of formulae, in order to visualise sets of cells with some commonality as a single cognitive unit. Our model encourages a structure in which such cognitive units are expressly defined. We also draw on the work of [Miller & Hermans, 2016] and in particular their notion of a "semantic axis" which has parallels with the way we use our template cells.

In tackling these issues at the level of the underlying data model – our grid replacement – we are proposing an organising principle for spreadsheet-like data, rather than a blueprint for a tool as such. With the right choice of model, we can make certain common spreadsheet structures (e.g. multi-level tables, and families of similar tables) emerge naturally, as opposed to requiring new abstractions. So ours is a complementary approach to the tools mentioned above, some of which might sit more comfortably upon this data model than upon the traditional spreadsheet grid.

We have implemented our data model, called the "lish", and its associated operations in a small demonstration application. It is work in progress but currently supports basic editing and interactive transformation of spreadsheet data into lish form, though not (as yet) calculation. It includes a layout engine that forms the basis of a fledgling GUI and which was used to produce the screenshots in Figures 2.2 to 2.4. It reads and writes data as JSON.

In the next section, we explain what our new data model consists of by means of some example spreadsheets, and in section 3 we summarise the underpinning theory. Section 4 wraps up with our planning for the next stages of this research.

## 2. AN ALTERNATIVE TO THE SPREADSHEET GRID

### 2.1 Row and Column Headers are Part of the Worksheet

Figure 2.1 shows a small example spreadsheet (produced in LibreOffice) which takes some population and geographical area data and calculates population density by region. The cells in the two population density columns each contain a formula which references the area and the relevant annual population column. In the figure, one such formula is visible in cell E8.

We begin our reformulation of the spreadsheet grid with a very simple, almost trivial change. It concerns the row and column headers: those marginal "cells" (see Figure 2.1) containing the row numbers and column letters respectively. In a normal spreadsheet they form a labelling area and are not really proper "cells", but our first change is to make them so. We allow the user to select and edit them, just like any other cell. The expected (though not mandated) usage is that the user will position column headings in these marginal areas, rather than in the internal cells of the worksheet.

In a normal spreadsheet, the user may set properties (such as numeric format) across an entire row or column by clicking in the appropriate header. Our equivalent is to allow the user to set properties on the equivalent marginal cell, whereupon they are inherited by the corresponding row, or column.

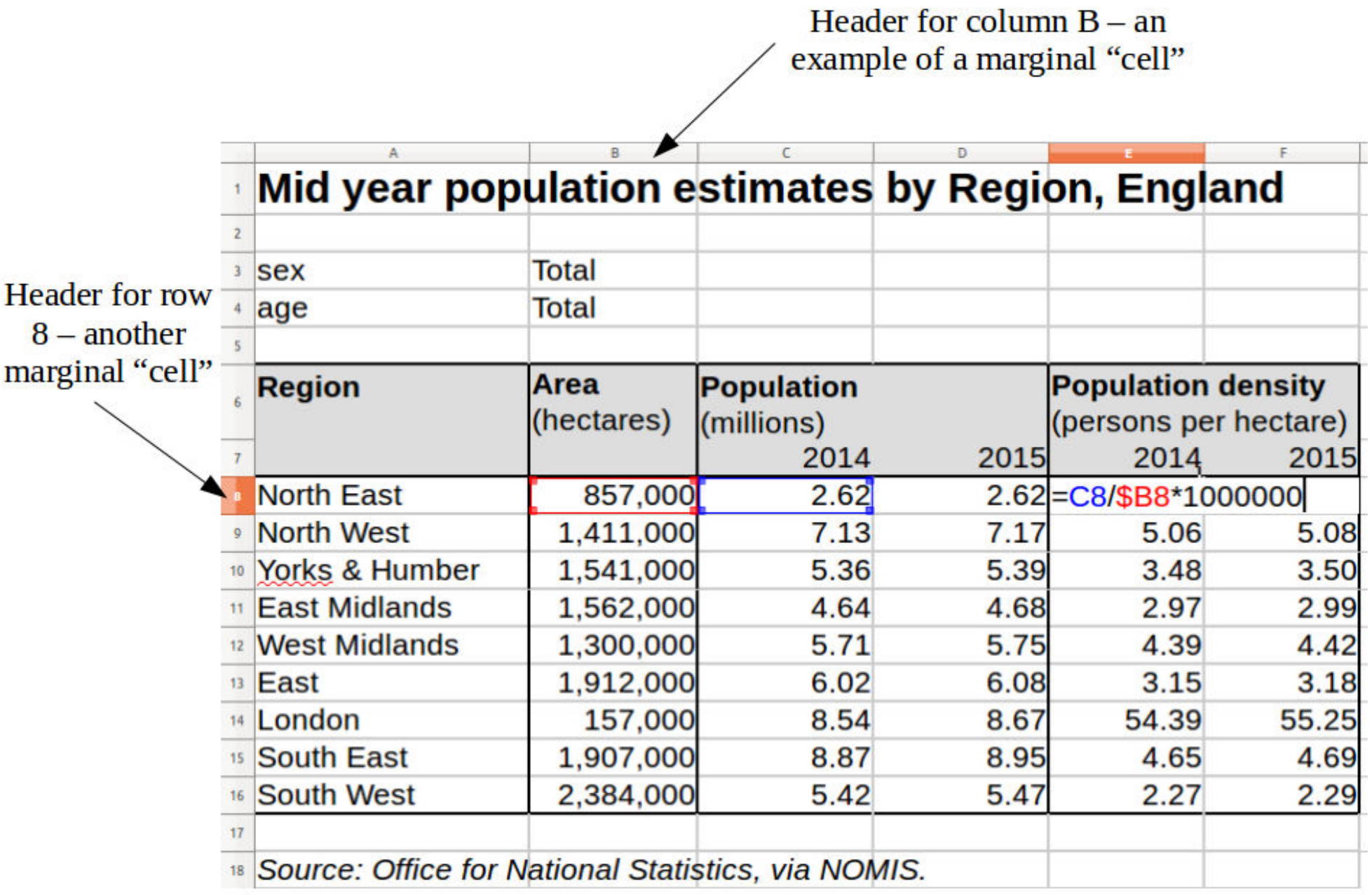

Mid year population estimates by Region, England

| | A | B | C | D | E | F |
|---|---|---|---|---|---|---|
| 1 | Mid year population estimates by Region, England | | | | | |
| 2 | | | | | | |
| 3 | sex | Total | | | | |
| 4 | age | Total | | | | |
| 5 | | | | | | |
| 6 | Region | Area (hectares) | Population (millions) | | Population density (persons per hectare) | |
| 7 | | | 2014 | 2015 | 2014 | 2015 |
| 8 | North East | 857,000 | 2.62 | 2.62 | =C8/$B8*1000000 | |
| 9 | North West | 1,411,000 | 7.13 | 7.17 | 5.06 | 5.08 |
| 10 | Yorks & Humber | 1,541,000 | 5.36 | 5.39 | 3.48 | 3.50 |
| 11 | East Midlands | 1,562,000 | 4.64 | 4.68 | 2.97 | 2.99 |
| 12 | West Midlands | 1,300,000 | 5.71 | 5.75 | 4.39 | 4.42 |
| 13 | East | 1,912,000 | 6.02 | 6.08 | 3.15 | 3.18 |
| 14 | London | 157,000 | 8.54 | 8.67 | 54.39 | 55.25 |
| 15 | South East | 1,907,000 | 8.87 | 8.95 | 4.65 | 4.69 |
| 16 | South West | 2,384,000 | 5.42 | 5.47 | 2.27 | 2.29 |
| 17 | | | | | | |
| 18 | Source: Office for National Statistics, via NOMIS. | | | | | |

Figure 2.1. A spreadsheet which calculates population density by geographical region.

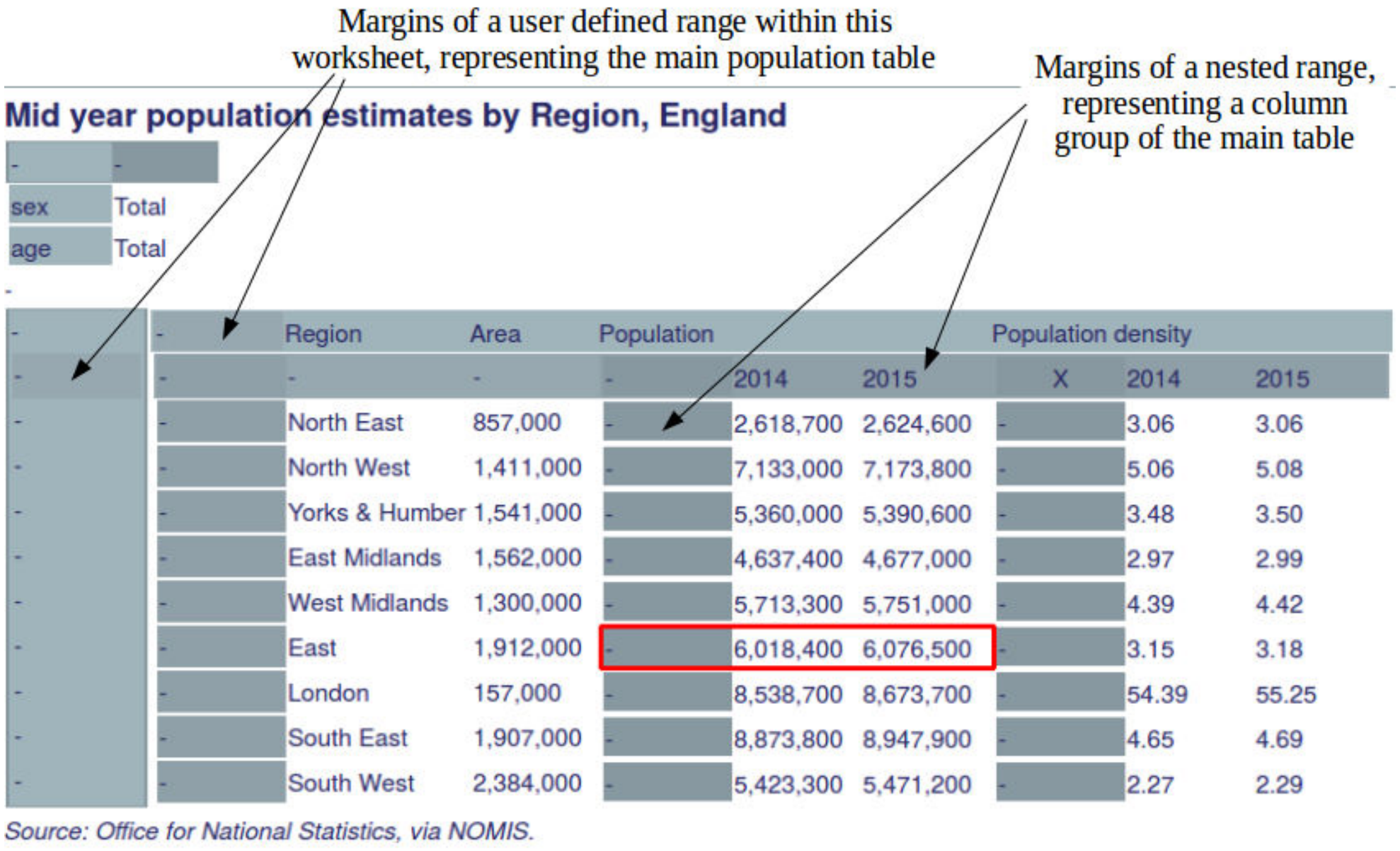

Mid year population estimates by Region, England

| | |
|---|---|
| - | - |
| sex | Total |
| age | Total |
| - | |

| | | | | | | | | | |
|---|---|---|---|---|---|---|---|---|---|
| - | - | Region | Area | Population | | | Population density | | |
| - | - | - | - | - | 2014 | 2015 | X | 2014 | 2015 |
| - | - | North East | 857,000 | - | 2,618,700 | 2,624,600 | - | 3.06 | 3.06 |
| - | - | North West | 1,411,000 | - | 7,133,000 | 7,173,800 | - | 5.06 | 5.08 |
| - | - | Yorks & Humber | 1,541,000 | - | 5,360,000 | 5,390,600 | - | 3.48 | 3.50 |
| - | - | East Midlands | 1,562,000 | - | 4,637,400 | 4,677,000 | - | 2.97 | 2.99 |
| - | - | West Midlands | 1,300,000 | - | 5,713,300 | 5,751,000 | - | 4.39 | 4.42 |
| - | - | East | 1,912,000 | - | 6,018,400 | 6,076,500 | - | 3.15 | 3.18 |
| - | - | London | 157,000 | - | 8,538,700 | 8,673,700 | - | 54.39 | 55.25 |
| - | - | South East | 1,907,000 | - | 8,873,800 | 8,947,900 | - | 4.65 | 4.69 |
| - | - | South West | 2,384,000 | - | 5,423,300 | 5,471,200 | - | 2.27 | 2.29 |

*Source: Office for National Statistics, via NOMIS.*

Figure 2.2. An alternative representation of the population data, in which internal margins have been defined. (For reasons of space, the very outermost margins have been omitted.)

## 2.2 Internal and Nested Margins

We pursue this idea a little further by allowing the user to define further ranges within the worksheets – somewhat like conventional named ranges, but in our model the name is optional. Each of these ranges now acquires its own margins, which behave like those of the global worksheet described above but apply only to their local range. Figure 2.2 shows our alternative representation of the data above, in which a number of such ranges have been defined: there is one for the small independent table at the top (the metadata giving the scope with regard to age and sex), and a larger one for the main table. The example also shows a further extension to this scheme, namely that ranges may be nested inside one another. The main table is in this way partitioned into three internal sub-tables: the left hand one contains the Region and Area columns, the middle one contains Population and the right hand one, Population Density. The shaded cells are all marginal, in the sense described above, and the nesting is visually apparent from the double margins on the main table – one margin for the top level table, and one for each nested range within it.

### 2.3 Cursor Behaviour

It will be noticed that in Figure 2.2, three cells are selected by a red wireframe cursor: the population figures for the two years for "East" region, and the corresponding internal marginal cell. However this is not an arbitrary selection of the kind one would obtain by dragging across a range of cells in a conventional spreadsheet. Rather, the selection cursor in our scheme is allowed to select any of the ranges that we set up above in their entirety, or any row, column or (of course) single cell within such a range. Pressing the up-arrow in our demo application moves the cursor to the corresponding three cells (including one marginal) for "West Midlands" – it does not collapse the selection back to a single cell. The user can drill in or out of the ranges they have defined and hence navigate at varying granularity.

Our rationale for this behaviour is that when building a spreadsheet model, users often wish to perform some operation uniformly upon a selection of cells – but not just any selection. The kinds of selection that are needed very often correspond to one of the ranges which a judicious application of this scheme would generate, or a row or column within one of those ranges. Allowing the cursor to behave in this way facilitates forming such cell selections and reduces the risk of "out-by-one" errors that an imprecise mouse movement might cause.

### 2.4 Marginal Formulae

Just as the inputs to a formula often consist of the kinds of cursor selections we have just described, so too do the outputs. That is, in a spreadsheet we often have a range of cells containing a formula that is identical except for some regularly updating pattern in the cells referenced. This formula is only required to be located in one place; we propose that that place should be the marginal cell which governs the range in which its results are to appear. This is similar in concept to the array formulae of traditional spreadsheets but more flexible, as it is perfectly legal for insertions or deletions of cells to take place within the output range, and the idea generalises to more complex structures as we shall shortly explain.

As an example, in Figure 2.2 we could locate column formulae for the two columns calculating population density (from population count and hectares) in the top marginal cells of their respective columns, assuming that the marginal labels are separable from the formulae. Better yet, we might observe that this formula is actually the same in both columns and enter it only once, in the cell marked "X" in the figure, which forms the pivot of the population density sub-table. Note that the marginal cells themselves are to be excluded from any arithmetic (e.g. column sums) so may be included in a selection with impunity.

In a strong version of this scheme, we might even go so far as to constrain the user to place formulae only in marginal cells. For any formula that is a one-off, yielding a single scalar value, the user would be required to create a mini "table" containing one marginal cell to contain the formula (and optionally, a label) and one ordinary cell to contain the answer. A weaker version is to allow individual cells to override their parent margin, with a warning indicator when this is taking place. Similarly, a column formula will override a range formula – the "innermost" one wins – also with a warning. In the strong version, we would also forbid arbitrary cell selections: the only multiple selections would be those corresponding to some margin. This would greatly reduce the risk of inconsistent formulae, though clearly with a trade-off against flexibility.

### 2.5 Margins as Templates: Generalising to Higher Dimensions

In order to follow our scheme to its logical conclusion, we will now make a shift in perspective: we will continue to deal with internal margins and with ranges of the worksheet, but we will view our margins as *templates* for the body of the range which they govern. The header row of a table, after all, is providing a description for the rows that follow: it says how many cells are to be in each row, i.e. how many columns there are, and by providing labels it tells the user what the meaning of those cells is. There are fewer use cases for the left hand margin performing a similar role, but in principle it could do so: for example, a cell in this margin could specify a highlight property to be applied to the entire row, and that too could be viewed as a form of template where the first cell acts as a basis for the others. A null template – i.e. an empty cell in the margin – may be interpreted as imposing no prior expectation on what the subsequent cells can contain.

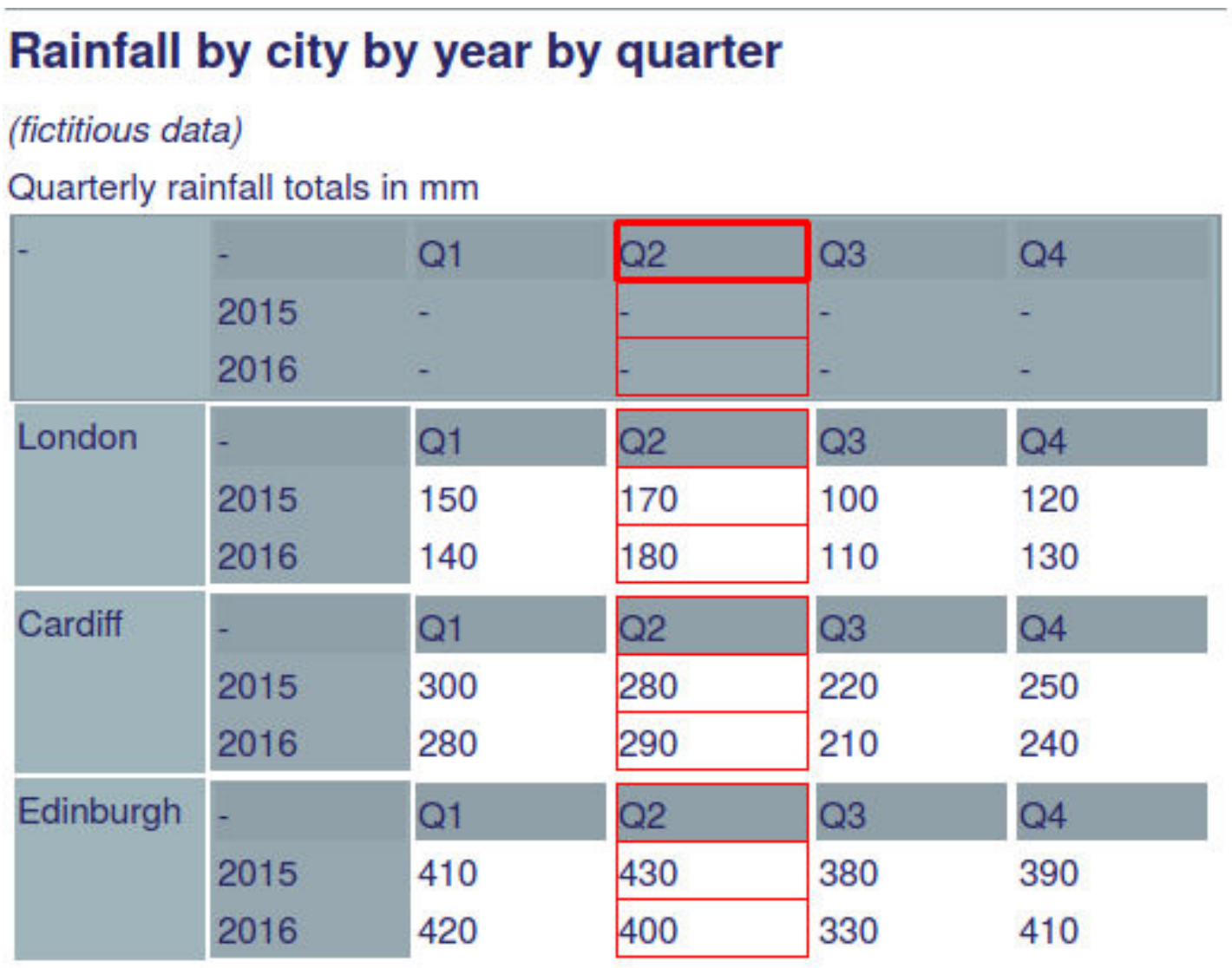

**Rainfall by city by year by quarter**

*(fictitious data)*

Quarterly rainfall totals in mm

| - | - | Q1 | Q2 | Q3 | Q4 |
|---|---|---|---|---|---|
| | 2015 | - | - | - | - |
| | 2016 | - | - | - | - |
| London | - | Q1 | Q2 | Q3 | Q4 |
| | 2015 | 150 | 170 | 100 | 120 |
| | 2016 | 140 | 180 | 110 | 130 |
| Cardiff | - | Q1 | Q2 | Q3 | Q4 |
| | 2015 | 300 | 280 | 220 | 250 |
| | 2016 | 280 | 290 | 210 | 240 |
| Edinburgh | - | Q1 | Q2 | Q3 | Q4 |
| | 2015 | 410 | 430 | 380 | 390 |
| | 2016 | 420 | 400 | 330 | 410 |

Figure 2.3. Internal structure in a margin defining a three dimensional array. All the grey cells are marginal, with the level of shading denoting depth within the structure.

We have stated that marginal cells are to be editable just like any other cell. So, if we can define ranges anywhere within the sheet, what would be the implication of defining them within the margins themselves? From the template perspective, the answer is that the ranges governed by those margins must conform to whatever structure the margins define.

One application of this pattern is the formation of three dimensional arrays. Figure 2.3

shows some rainfall data broken down by quarter, year and city. The top band of shaded cells is itself a two-dimensional table. Since this band is marginal it contains no actual data but rather acts as a template, forming the base plane for a stack of tables of the same structure. Hence, we have our three dimensional array. In the figure, the user has positioned the cursor at cell level in the “Q2” heading of the template. Our application has automatically highlighted a secondary selection comprising all those cells which are governed by this marginal cell. Had the user selected the “Q2” in the internal margin belonging to Cardiff, only the two rainfall values relating to Cardiff in that column would have appeared in the secondary selection. Recall that marginal cells are omitted from calculations, so this selection could be used as the input to a formula producing summary statistics on rainfall.

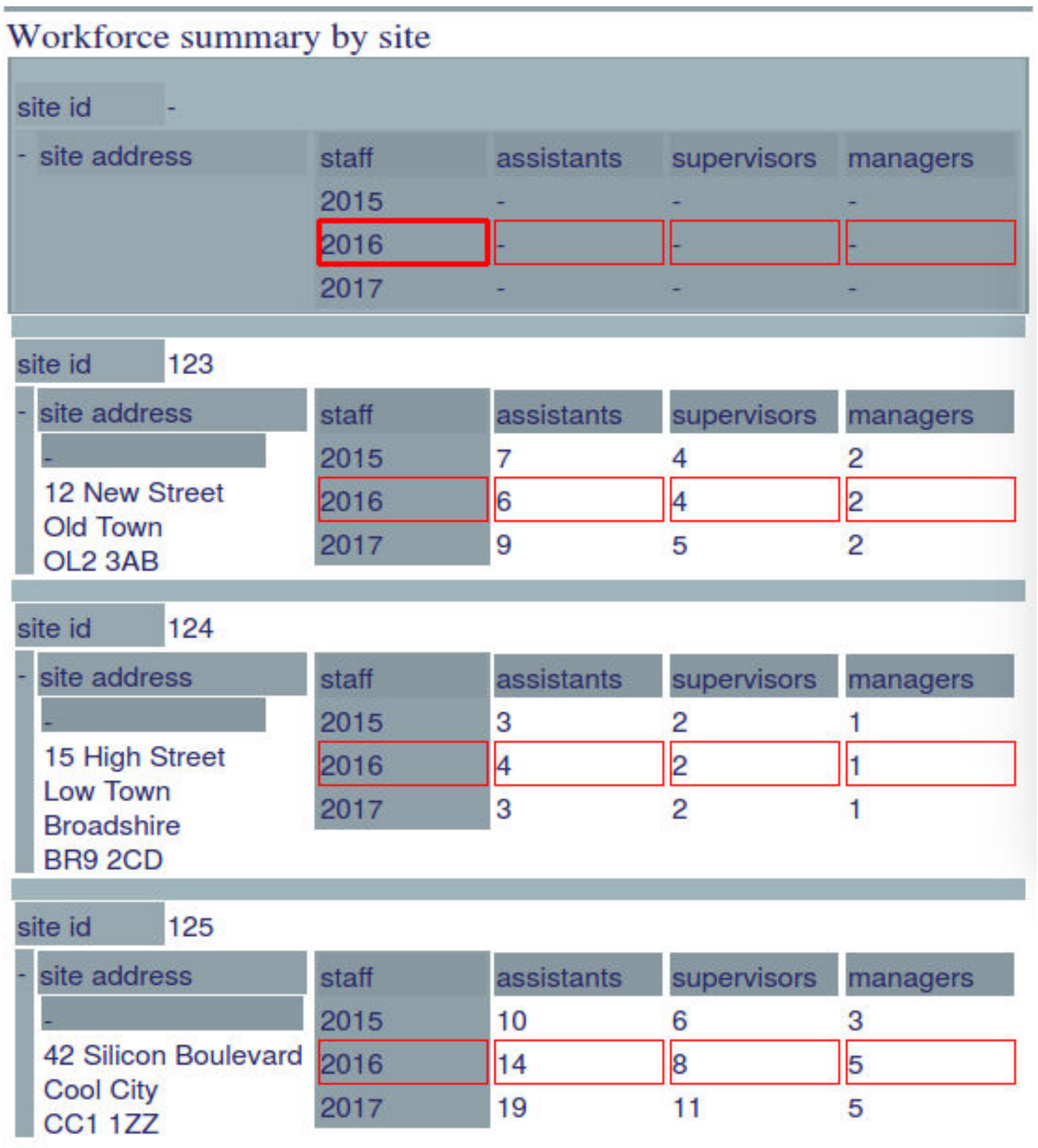

Workforce summary by site

| site id | - |
|---|---|

| - site address | staff | assistants | supervisors | managers |
|---|---|---|---|---|
| | 2015 | - | - | - |
| | 2016 | - | - | - |
| | 2017 | - | - | - |

| site id | 123 |
|---|---|

| - site address | staff | assistants | supervisors | managers |
|---|---|---|---|---|
| - | 2015 | 7 | 4 | 2 |
| 12 New Street | 2016 | 6 | 4 | 2 |
| Old Town | 2017 | 9 | 5 | 2 |
| OL2 3AB | | | | |

| site id | 124 |
|---|---|

| - site address | staff | assistants | supervisors | managers |
|---|---|---|---|---|
| - | 2015 | 3 | 2 | 1 |
| 15 High Street | 2016 | 4 | 2 | 1 |
| Low Town | 2017 | 3 | 2 | 1 |
| Broadshire | | | | |
| BR9 2CD | | | | |

| site id | 125 |
|---|---|

| - site address | staff | assistants | supervisors | managers |
|---|---|---|---|---|
| - | 2015 | 10 | 6 | 3 |
| 42 Silicon Boulevard | 2016 | 14 | 8 | 5 |
| Cool City | 2017 | 19 | 11 | 5 |
| CC1 1ZZ | | | | |

Figure 2.4. A family of objects, with the template “2016” cell selected and a cross-cutting secondary selection spanning the entire family. The marginal cell above each site address arises because a single cell in the template is replaced by a range in that site instance.

Figure 2.4 shows a similar three dimensional example, but this time we do not have a “pure” three dimensional array but rather a family of objects, each containing an assortment of fields including a two dimensional table. The data are from a spreadsheet that a (fictitious) company might use as part of workforce planning. For a number of sites, the sheet shows a site id, site address, and a table of headcount broken down by grade and year.

The margin-as-template principle lends itself easily to this pattern. The object format is defined in the top shaded band of cells, and this governs a range of objects each conforming to that format. In the figure, the user has placed the cursor at the “2016” cell of the “staff” table. As in the previous example, the system has automatically generated a secondary selection comprising the cells governed by this margin. This time, however, the secondary cells are not all adjacent – we have a cross-cutting selection containing the staffing figures for all sites and all grades, for the year 2016 only.

The data presented in this example could alternatively be held in a normalised database. The structure we have presented here captures some of the discipline associated with that option while presenting the data in a more human readable form. The marginal cells allow some of the cross-cutting selections that would otherwise require a separate query or pivot table to be generated on the data in-place, in a visually intuitive way.

Higher dimensional structures may be obtained if necessary by a generalisation of the same procedure: the outermost margin is set up as a table of tables of tables, to whatever depth is required.

**2.6 Dynamic Cell Allocation**

Careful examination of Figure 2.4 will reveal another feature of the way we interpret template elements: a single cell within a template is allowed to spawn an embedded range of cells at the corresponding position within the range that it governs. The site address is an example of this. In the template, only a single cell represents the site address. But for each individual instance of a site, a one-dimensional range of cells has been embedded at that location in order to accommodate an address with a variable number of lines. Our layout engine is conversant with this behaviour and accommodates the variable object size appropriately, while maintaining alignment over elements that form part of some larger table.

Our wider intention in defining this behaviour is to allow formulae to return results whose number of cells is not known at design-time. Hence for future work, we propose this as an approach to addressing the dynamic allocation problem for spreadsheets.

**2.7 Dispensing with the Global Grid**

We claimed in the introduction that our data model would be an alternative to the normal spreadsheet grid, as opposed to an additional construct to be overlaid upon it. We have framed the discussion above in terms of "ranges" of cells selected over the normal grid, as that is the way one would go about refactoring an existing spreadsheet into our form. But once we have it in this form, we can indeed dispense with the concept of an underlying grid altogether. All that we require is the nested system of cell ranges with their margins. Similarly, if starting from a blank document, we could create the various templates and enter the data without ever going via a global underlying grid at all.

We note here that eliminating the global grid has cleared up a minor visual annoyance with the normal spreadsheet: when two tables are vertically juxtaposed on the same worksheet, they are forced to share common column widths, even though the content in some column of the upper table might be suited by a very different width to the corresponding column in the lower one. In Figure 2.2, our layout engine has recognised that the small table at the top is a separate range from the main table, and has broken this unwanted coupling.

## 3. THE "LISH" DATA MODEL

This section discusses the theory of the underlying data model needed to support the behaviour described above. Formal specification is deferred to a future more technical document.

At first glance, it might seem as if what we have are simply cells within cells – a normal spreadsheet grid in which some cells – we might call them "super-cells" – are allowed to contain other cells. To a first approximation this (rather obvious) extension to the standard spreadsheet is indeed what we propose, but it's not quite the whole story. The problem is that we want some of the "super-cells" to contain grids that are the same size and in alignment with their neighbours, or even with other "super-cells" elsewhere on the sheet entirely, whereas others are under no such constraint. We would like a simple way to capture that without introducing a spaghetti-network of explicit links.

The model that we propose instead is called the "lish". (The name is a portmanteau of "list" and "hash table" since in its fully specified form it shares some characteristics of both; it is the list-like properties that we focus on in this paper.) A lish is a list of elements, each of which can be either a single atomic value (numeric, string or boolean) or a further lish. The atomic values correspond to individual cells and the sublists to the ranges as described in the previous section. For a list to be a valid lish, it must meet certain criteria to be described shortly.

To develop the theory we will set aside the graphical world of the spreadsheet and fall back on a simpler textual notation. We could represent a small subset of the earlier population data using the following list of lists:

```
[[null, "Region", "Area"],
 [null, "North East", 857000],
 [null, "North West", 1411000],
 [null, "Yorks & Humber", 1541000]]
```

Following common programming languages, we have used here a notation where lists are enclosed in square brackets with their constituent elements separated by commas. A null denotes an empty cell; the four nulls above correspond to the cells in the left hand margin of the table in Figure 2.2. "Region" and "Area" are column names and hence occupy cells in the top margin. In this example, there are no row names, so the left hand margin contains only nulls.

We want the list of lists above to represent a table: the third element of each sublist, for instance, is related to the third element of every other sublist. Taken together, these elements form the Area column. But there could be other lists of lists where there is no tabular interpretation, and the third element in one list is not related in any particular way to the third element in any other. We need a way to distinguish these cases. This could be achieved by the use of markup tags (as in HTML) or validation against an external artefact (as in JSON Schema) but we prefer here to distinguish them by means of the list structure itself. Our rationale is that in a spreadsheet, the user creates a structure simply from the shape of the data as entered. If we can make the supporting structure self-describing, we avoid the ceremony of having to tell the system how it is supposed to be interpreted.

The previous section introduced the idea of marginal cells being regarded as "templates" for other cells in their range, and this is the approach we take to constructing the system of nested lists that is to represent them. Every list that is a *lish* must contain at least one element – there is no empty lish. This first element in each lish is used to represent a marginal cell. It does not contain normal data, but instead performs the template role for subsequent elements in the same lish.

The rule, stated informally, is that the subsequent elements must contain "at least as much structure" as their template, but are allowed to contain more. Everywhere the template contains an atom, the subsequent elements in that lish may contain either an atom or another lish; everywhere the template contains a lish, the subsequent elements must also contain a lish, and it must be of the same length.

For example, the following lists are valid lishes:

`[null, 1, 2, [3, 4], 5]`
– template is atomic, subsequent elements are mix of atomic and list

`[[null, null, null], [null, 1, 2], [null, 3, 4]]`
– template is a list of three, so are subsequent elements (a table)

But the following lists are not valid lishes:

`[[null, null], 1, 2, 3]`
– template is a list, but subsequent elements are atoms

`[[null, null], [1, 2, 3]]`
– template is a list of two, but next element is a list of three

The above rule is applied recursively, within both the template and the subsequent elements. So a template element of `[[null, null], null]` at the head of some lish would be illegal: the template now has a template of its own, a list of two, which the final null does not respect. `[null, [null, null]]` would be permissible since an atomic template may be succeeded by a list. Some more examples illustrate further the recursive application of the template rule:

`[[null, [null, null]], [1, [2, 3]], [4, [5, 6]]]`
– **legal**, all three elements of the outer lish have the same form
– and all are legal lishes internally

`[[null, [null, null]], [1, [2, 3]], [4, [5, [6, 7, 8]]]]`
– same as above except for the final `[6, 7, 8]`
– and **legal:** the `[6, 7, 8]` has an atomic template, so a lish is allowed here

`[null, [[1, 2], 3, 4]]`
– a null template followed by a lish is valid
– but **illegal**, since the atom, `3`, following `[1, 2]` is not

This rule enables us to define a number of commonly occurring structures in spreadsheets. For example, the inner top margin of the main table of Figure 2.2 is the lish `[null, null, null, [null, 2014, 2015], ["X", 2014, 2015]]`. This sets the template for subsequent rows of its table as three columns, followed by two column groups each containing a further three columns. A regular three-dimensional array is defined by a two-dimensional table within its template. For example, the rainfall data of Figure 2.3 contains within its top margin the lish:

```
[
 [null, 2015, 2016],
 ["Q1", null, null],
 ["Q2", null, null],
 ["Q3", null, null],
```

```
 ["Q4", null, null]
]
```

Since the template position may contain any lish, we are not restricted to regular arrays. For example, the template for the workforce data of Figure 2.4 is:

```
[null,
  ["site id", null],
  [null,
    ["site address", " "],
    [["staff", 2015, 2016, 2017],
     ["assistants", null, null, null],
     ["supervisors", null, null, null],
     ["managers", null, null, null]]
  ]
]
```

The lish itself is agnostic to whether tables are represented row-wise or column-wise. By default, the layout engine simply alternates the orientation of sublists by depth of nesting, but this may be overridden on specific lishes according to user preference.

Of course, from the user's point of view all this is done visually by selecting ranges, which then acquire margins, and embedding further ranges within those margins when more complex templates are required: the user is insulated from the nested bracket notation entirely. But underneath, ranges are lists, individual cells are atoms, first elements of lists are marginal cells, and the template rule as stated ensures that structures respect their template while allowing additional embedded structure (new sublists) to be added over and above that template. The dynamic allocation behaviour noted in section 2.6 arises because it is legal to associate an atom in a template with a list in a cell that it governs. We use an algorithm that parses the lish structure in order to associate each marginal cell with the appropriate set of cells within the table body, yielding the "secondary selections" of Figures 2.3 and 2.4. The visual layout engine uses a similar algorithm (though not identical, since its output is always a two dimensional projection) to determine which cells are in equivalent rows or columns to others and so need to be kept aligned.

It should be noted that "atomic" elements within the template can actually perform at least three roles: they can contain labels; they can define formulae; and they can specify formatting information. So they are in fact more complex objects than the simple textual representation above would imply, but they are regarded as atomic with respect to their role in defining further structure.

There is a caveat when there are nested template elements: it can sometimes arise that a subsequent element is apparently governed simultaneously by two or more conflicting templates. A typical use case would be when we wish to group both rows and columns within a table. An extension to the theory, outside the scope of this paper, accommodates that situation by "composing" the rival templates: briefly, the element under contention acquires structure from all its templates combined, but without duplicating any structure that had a common origin. The mathematics of specifying the lish becomes rather more involved as a result, but the intuition that the first element of each lish governs the form of its successors holds good. The algorithms used in the prototype application accommodate such cases.

## 4. DISCUSSION

### 4.1 The Lish in Summary

We have proposed an alternative data model, the "lish", that could be used as an underlying representation for a spreadsheet, in place of the traditional grid of cells. The model is based on nested lists of cells, with a constraint that enables certain structures that encompass parallel lists to be captured.

The model maps to a user's point of view in which marginal cells acquire the same capabilities as any other cell, and behave like templates over those ranges of cells which they govern. Moreover the user may define internal ranges, with their own internal marginal cells, and may even define these within the margins themselves giving rise to more complex structures.

We have shown how from the simple ground rules upon which the lish is defined, a number of patterns frequently arising in spreadsheets (including tables with grouped rows and columns, higher dimensional tables, and families of similar objects) can be made to emerge. The nested structure allows these objects to be navigated at multiple granularity, and the association between marginal cells and sub-ranges allows formulae to be defined in a way that minimises replication: the spreadsheet can be made DRY.

### 4.2 The User Perspective

The benefits of having a lish structure once it is in place are clear enough: users can work at multiple granularities, easily select and manipulate logically coherent portions of the worksheet, and avoid both the tedium and the risks associated with manually replicating tables and formulae. But if users are to gain these advantages, they must be persuaded of the merits not only of learning to use a lish representation (though that would be true of any new system) but also of investing the effort in applying the extra structure to their data by means of defining ranges and associated templates. Will they perceive it as worth their effort to get there?

A full discussion of this question must await the outcome of the user studies (see next subsection). But we have already taken some steps to smooth the path for users as much as possible. In the spirit of the traditional spreadsheet, we do not impose a large number of new concepts upon the user: the large variety of structures that may be formed all arise from two basic ones, the "ranges" and "templates". The former will be familiar to anyone who has already worked with named ranges, and the latter are not conceptually distant from the usual working pattern of designing a small part of the spreadsheet and then copy-pasting to replicate it. Working with template cells feels very much like working with normal cells – they are not a separate type of object.

Another way we stay in the spirit of a normal spreadsheet is that structures are formed "by demonstration" from the shape of the data. Users are not forced to specify their application up front before the system will allow them to do anything. They can start from a regular grid (which of course is itself a perfectly valid lish) and apply the more detailed structure upon noticing which parts of their data belong together and would be worth enclosing in a range, and where there are repeating patterns that would be worth recycling as templates.

In future work, we propose to make this process of refining the structure more attractive to the user, by aligning it as closely as possible with the secondary notation that they might well wish to apply anyway. Referring back to Figure 2.1, the user has clearly decided to demarcate the main table by enclosing it in a box, to identify the column headings by shading them, and mark certain columns off into groups by means of vertical lines. These activities are at least closely parallel to the ones that need to be applied in order to create a lish structure, and we propose to exploit that parallel – perhaps even to the point that explicit formatting becomes deprecated in favour of a complete concordance between format and structure. An analogy might be the use of named styles in a word processor to make explicit the link between visual appearance and document structure, as opposed to having the user apply formatting piecemeal.

One further consideration for the user is conservation of complexity: merely expressing a complex structure in lish form does not simplify it. What it does do, however, is take certain constraints that in an ordinary spreadsheet would be implicit, or exist only in the user's head, and crystallise them. For example, if several tables are all supposed to perform the same type of calculation then this is a feature of the specification that always exists, whether the data are in lish form or not. But if they are, this part of the specification has been baked in and will be more robust to future maintenance. So although the complexity has not been removed, its associated risks have been mitigated.

### 4.3 Planned Future Work

A small prototype application is at an early stage of development. It is capable of carrying out some basic data editing and refactoring tasks interactively, but does not yet support actual calculations. We do not yet support separate worksheet tabs, but if implemented these will be treated as a visual presentation device only, so that the entire workbook is one lish. This would enable template tabs to control groups of further tabs without modification to the theory of section 3.

We are planning two trials with live users. The first will be to assess a range of real-world spreadsheet applications against the lish to see whether the intended benefits (especially around the capture of repeated structures and formulae) can indeed be realised. The second trial will take place once the prototype has been sufficiently developed to support the entire workflow of building a spreadsheet application from scratch. In this trial, users will complete tasks using the prototype with data where no lish structure has initially been applied. This will assess whether they are able to grasp the concepts of ranges and templates and apply them correctly so as to reformulate initially unstructured data into lish form.

The prototype has no support as yet for actually calculating anything. The calculation model will be formula based, as in the spreadsheet, but needs to operate not only on individual cells but at aggregate level. A “lish calculus” will therefore be specified. This will draw heavily on the use of vectorised calculations, as used in the R programming language [R Core Team, 2017], but will extend this concept to work with a general, nested, lish structure as opposed to only a one-dimensional vector or flat matrix.

## ACKNOWLEDGEMENTS

The authors would like to thank the Government Operational Research Service (UK) for supporting this research.